\documentclass[letterpaper]{aa}
\usepackage{graphicx}
\usepackage{txfonts}
\usepackage{natbib}
\usepackage{amssymb}
\newcommand{\lya}{Ly$\alpha$}
\newcommand{\ergscm}{erg\,s$^{-1}$\,cm$^{-2}$}
\newcommand{\kms}{km\,s$^{-1}$}

\begin{document}
   \title{Discovery of six \lya\ emitters near a radio galaxy at
   $z \sim 5.2$}

%   \subtitle{}

   \author{B.\ P.\ Venemans \inst{1} \and H.\ J.\ A.\ R\"ottgering \inst{1}
           \and R.\ A.\ Overzier \inst{1} \and G.\ K.\ Miley \inst{1}
           \and C.\ De Breuck \inst{2} \and J.\ D.\ Kurk \inst{3} \and
           W.\ van Breugel \inst{4} \and C.\ L.\ Carilli \inst{5} \and
           H.\ Ford \inst{6} \and T.\ Heckman \inst{6} \and P.\
           McCarthy \inst{7} \and L.\ Pentericci \inst{8}}

   \offprints{}

   \institute{Sterrewacht Leiden, P.O. Box 9513, 2300 RA, Leiden, 
              The Netherlands               
         \and
              European Southern Observatory, Karl Schwarzschild
              Stra{\ss}e 2, D-85748 Garching, Germany
         \and
              INAF, Osservatorio Astrofisico di Arcetri, Largo Enrico
              Fermi 5, 50125, Firenze, Italy
         \and
              Lawrence Livermore National Laboratory, 
              P.O. Box 808, Livermore CA, 94550, USA
         \and
              National Radio Astronomy Observatory, P.O. Box 0,
              Socorro, NM 87801, USA
         \and
              Dept. of Physics \& Astronomy, The Johns Hopkins University,
              3400 North Charles Street, Baltimore MD, 21218-2686, USA
         \and
              The Observatories of the Carnegie Institution of Washington,
              813 Santa Barbara Street, Pasadena CA, 91101, USA
         \and
              Dipartimento di Fisica, Universit\`a degli studi Roma Tre,
              Via della Vasca Navale 84, Roma, 00146, Italy
             }

   \date{Received May 10, 2004; accepted July 17, 2004}

   \abstract{We present the results of narrow-band and broad-band
   imaging with the Very Large Telescope\thanks{Based on
   observations carried out at the European Southern Observatory,
   Paranal, Chile, programs LP167.A-0409 and 70.A-0589} of the field
   surrounding the radio galaxy TN J0924--2201 at $z = 5.2$. Fourteen
   candidate \lya\ emitters with a rest-frame equivalent width of $>$
   20 \AA\ were detected. Spectroscopy of 8 of these objects showed
   that 6 have redshifts similar to that of the radio galaxy. The
   density of emitters at the redshift of the radio galaxy is
   estimated to be a factor 1.5--6.2 higher than in the field, and
   comparable to the density of \lya\ emitters in radio galaxy
   protoclusters at $z = 4.1, 3.1$ and 2.2. The \lya\ emitters near TN
   J0924--2201 could therefore be part of a structure that will evolve
   into a massive cluster. These observations confirm that substantial
   clustering of \lya\ emitters occurs at $z > 5$ and support the
   idea that radio galaxies pinpoint high density regions in the early
   Universe.

    }

   \maketitle
%
%________________________________________________________________

\section{Introduction}

One of the most intriguing questions in modern astrophysics concerns
the formation of structure in the early Universe
\citep[e.g.][]{bah97}. The narrow-band imaging technique can
efficiently select objects with a strong \lya\ line in a narrow
redshift range, and is therefore ideal for finding and investigating
overdense regions at high redshift \citep{ste00,mol01,shi03,pal04}.
For example, \citet{ste00} used narrow-band imaging to map the extent
of a large-scale structure at $z \sim 3.09$, discovered in a survey
for continuum-selected Lyman-break galaxies.  \citet{shi03}
serendipitously found a large-scale structure at $z \sim 4.9$ while
searching for \lya\ emitters in the Subaru Deep Field. Their results
demonstrate that Mpc-scaled structures have already formed by $z \sim
4.9$ and that \lya\ emitters must be very biased tracers of mass in
the early Universe.

Narrow-band imaging of distant powerful radio galaxies at $z = 2-4$
has shown that these objects are often located in rich environments,
possibly the early stages in the formation of massive clusters
\citep{pas96,lef96,pen00a,ven02,ven03b,kur04}. An interesting question
is out to which redshift such large-scale structures (protoclusters)
can be detected. The most distant known radio galaxy is TN
J0924--2201, with a redshift of $z = 5.2$ \citep{bre99}.
In this Letter, we describe broad- and narrow-band observations of
this radio galaxy, and report the discovery of 6 \lya\ emitters in the
field whose redshifts are close to that of the radio galaxy.
\footnote{Throughout this Letter, magnitudes are in the AB system and
a $\Lambda$-dominated cosmology with H$_0 = 65$ km s$^{-1}$
Mpc$^{-1}$, $\Omega_{M} = 0.3$, and $\Omega_{\Lambda} = 0.7$ is
assumed.}

\section{Observations and candidate selection}

\subsection{Imaging observations and candidate selection}

To search for candidate \lya\ emitters near TN J0924--2201,
narrow-band and broad-band ($I$- and $V$-band) imaging of the field
were carried out during two separate observing sessions in 2002 March
and April with the VLT Yepun (UT4), using the FOcal Reducer/ low
dispersion Spectrograph 2 (FORS2). The custom made narrow-band filter
had a FWHM of 89 \AA\ and a central wavelength of 7528 \AA, which
encompasses the wavelength of the \lya\ emission line at $z \sim
5.2$. The effective exposure times are 36860 s (narrow-band), 9750 s
($I$-band) and 3600 s ($V$-band). The seeing in the narrow-band,
$I$-band and $V$-band images is 0\farcs8, 0\farcs8 and 1\farcs05
respectively. The 3 $\sigma$ limiting magnitudes in an aperture with a
2\farcs0 diameter are 26.29, 26.65 and 26.80 for the narrow-band, $I$
and $V$-band respectively. For a \lya\ emitter at $z \sim 5.2$ with
negligible continuum, the 5 $\sigma$ limiting line luminosity is
$L_{\mathrm{lim}}(\mathrm{Ly}\alpha) = 3 \times 10^{42}$
erg\,s$^{-1}$. The images have an area useful for detecting \lya\
emitters of 46.8 $\square^\prime$.

A total of 3471 objects were detected in the narrow-band image with a
signal-to-noise greater than 5 using the program SExtractor
\citep{ber96}. For each object, the observed equivalent width was
calculated using a method that will be described in a future paper
(Venemans et al., in preparation). \lya\ emitters
at $z \sim 5.19$ with a rest-frame equivalent width of EW$_0 >$ 20
\AA\ would have an observed equivalent width (EW$_\lambda$) of 124
\AA. We find 24 such objects in the field. The $V$-band image was
used to identify low redshift interlopers with an emission line
falling in the narrow-band filter. Ten of the 24 objects with
EW$_{\lambda} > 124$ \AA\ were also detected in the $V$-band with a
signal-to-noise greater than 2, and had $V-I$ colors that were much
bluer ($V-I < 1.2$) than a $V-I$ color of $\sim 2.75$ as expected for
a galaxy at $z \sim 5.2$ \citep[e.g.][]{son04}. The remaining 14
candidates were our high priority candidates for follow-up
spectroscopy.

\subsection{Spectroscopy}
\label{spec}

For the spectroscopy, a mask was constructed which included the radio
galaxy and 8 of the 14 high priority candidate \lya\ emitters. This
was the maximum number that could be fitted on the mask. The rest of
the mask was filled with objects having an excess flux in the
narrow-band, but with a lower equivalent width than our selection
criterion and/or with a blue $V-I$ color. The observations were
carried out on 2003 March 3 and 4 using FORS2 on the VLT Yepun. The
mask was observed through the 600RI grism (with a peak efficiency of
87\%) with 1\farcs4 slits. The pixels were $2\times2$ binned to
decrease the readout time and noise, giving a spatial scale of
0\farcs25 pixel$^{-1}$ and a dispersion of 1.66 \AA\,pixel$^{-1}$. The
total exposure time was 20676 s. The mean airmass was 1.23 and the
seeing in the individual frames varied between 0\farcs7 and 1\farcs0,
giving a spectral resolution of 185--265 \kms\ for point sources. For
the wavelength calibration, exposures were taken of He, HgCd, Ar and
Ne lamps. The rms of the wavelength calibration was always better than
0.25 \AA\ ($\sim 10$ \kms).

\section{Results}

The radio galaxy and all of the 8 observed candidate \lya\ emitters
showed an emission line near 7500 \AA. The redshift of the radio
galaxy of $z = 5.1989 \pm 0.0006$ is consistent with the redshift of
$z=5.2$, reported by \citet{bre99}.

\begin{figure}
\resizebox{\hsize}{!}{\includegraphics{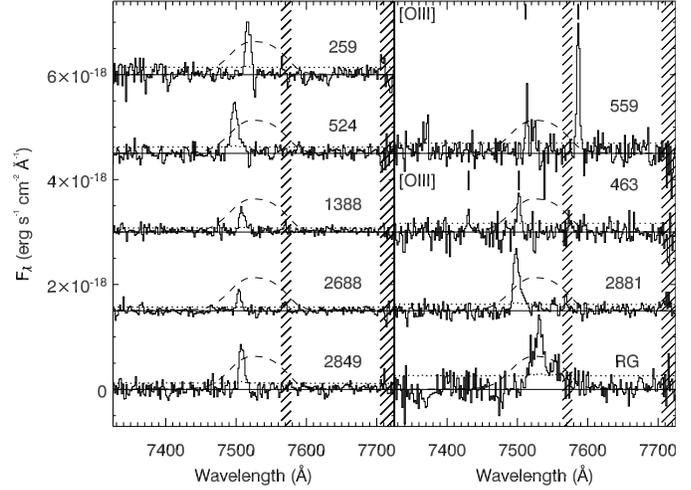}}
\caption{\label{spectra} Part of the spectra of the eight
  spectroscopically observed high priority emitters and the radio
  galaxy. For clarity the spectra are offset by $1.5 \times 10^{-18}$
  \ergscm. The solid lines indicate the zeropoint of the spectra, the
  dotted lines the $1 \sigma$ uncertainty in the data, and the dashed
  lines are the scaled transmission curves of the narrow-band
  filter. The regions in the spectrum where strong telluric skylines
  dominate are indicated with hashed lines.}
\end{figure}

Two of the 8 candidate \lya\ objects (emitters \# 463 and \# 559) are
identified with [\ion{O}{iii}] $\lambda 5007$ at $z \sim 0.5$,
confirmed by the accompanying lines [\ion{O}{iii}] $\lambda 4959$ and
H$\beta$ (Fig.\ \ref{spectra} and Table \ref{zlya}). The other six
spectra (Fig.\ \ref{spectra}) did not show any other emission line in
a wavelength range covering more than 3300 \AA\ (see Table
\ref{zlya}).

To distinguish high redshift \lya\ emitters from low redshift
interlopers various tests can be applied \citep[see][for a
review]{ster00}.

{\em Asymmetric line profile} A characteristic feature of a high
redshift \lya\ line is the flux decrement on the blue wing of the
\lya\ emission \citep[e.g.][]{daw02}. Following \citet{rho03}, the
asymmetry of an emission line can be described by the parameters
$a_\lambda$ and $a_f$. These parameters measure the ratio of the line
width and line flux redward and blueward of the line peak and depend
both on the characteristics of the line (line width, amount of
absorption, merged doublet) and on the resolution of the spectrum
\citep[][and reference therein]{rho03}. Simulations of observed
spectra indicate that Gaussian \lya\ lines with a FWHM of 150--800
\kms\ and with the blue side fully absorbed have $a_{\lambda} =
1.0-1.6$ and $a_f = 1.0-1.4$, while [\ion{O}{ii}] $\lambda 3727$
emitters have $a_{\lambda} \approx 0.9$ and $a_f = 0.8-0.9$. This is
consistent with values found by \citet{rho03}, who measure typical
values of $0.9 < a_f < 1.9$ and $0.9 < a_\lambda < 3.1$ for a sample
of high redshift \lya\ emitters and for [\ion{O}{ii}] emitters at $z
\sim 1$ $a_f \approx 0.8$ and $a_\lambda \approx 0.9$. Only two of the
emission lines (of emitters \# 1388 and \# 2881) have a
signal-to-noise that is high enough to measure the asymmetry. These
two lines are (marginally) asymmetric (with $a_\lambda = 2.0 \pm 0.9
(2.2 \pm 0.6)$ and $a_f = 1.7 \pm 0.8 (1.4 \pm 0.6)$ for emitter \#
1388 (2881)), an indication that emitters \# 1388 and \# 2881 are
\lya\ emitters at $z \sim 5.2$.

{\em Continuum break} A high redshift \lya\ emitter must have a
continuum break across the \lya\ line, caused by the \lya\ forest
between the galaxy and the observer. \citet{mad95} predict a break of
a factor $\sim 5$ across the \lya\ line at $z \sim 5$. To measure
continuum in our spectra, regions that were not effected by strong
telluric lines were chosen redward and blueward of the emission line.
Four spectra had a significant ($> 3\sigma$) detection of continuum emission
redward of the emission line, resulting in $2\sigma$ lower limits on
their flux decrements of 3.6 -- 5.3. Such large breaks in the optical
are exclusively found in high redshift objects
\citep[e.g.][]{ster00}. \citet{ham97} showed that observed [\ion{O}{ii}]
emitters at $0.5 < z < 1.0$ have %rest-frame colors of $U-V < 1.4$, and
%that objects with rest-frame $U-V < 1.4$ have 
a total 4000 \AA\ and Balmer
break of factor $< 3$. Therefore, the continuum break measured in four
of the emitters is most likely caused by neutral \ion{H}{i}
absorption, and hence these emitters can be identified with \lya\
emitters at $z \sim 5.2$.

{\em Equivalent width} The emission line objects have observed
equivalent widths in excess of $\sim 250$ \AA. The two emitters which
do not show a convincing line asymmetry and do not show continuum
both redward and blueward of the emission line, have observed
equivalent widths of EW$_\lambda > 540$ \AA. This would correspond to
a rest-frame equivalent width of $>$ 269 \AA\ if the emission line
is [\ion{O}{ii}] $\lambda 3727$ at $z \sim 1.0$. Such high
[\ion{O}{ii}] equivalent width emitters are rare. The total number of
$z \sim 1$ [\ion{O}{ii}] emitters expected in our field, derived from
\citet{tep03}, is $\sim 1$. However, the fraction of [\ion{O}{ii}]
emitters with a rest-frame EW $>$ 200 \AA\ is $< 2.5$\% \citep{tep03},
which indicates that these two emission line objects are probably not
[\ion{O}{ii}] emitters at $z \sim 1$, but \lya\ emitters at $z = 5.2$.

{\em Emission line ratios} As mentioned above, no other emission lines
were found in the spectra of the emitters. To estimate the likelihood
that the emitters are [\ion{O}{ii}] emitters, we can derive an upper
limit on the flux of the [\ion{Ne}{iii}] $\lambda 3869$ line and
compare that to local emission line galaxies \citep{fyn01}. Stacking
the six spectra to increase the signal-to-noise, we find an upper
limit on the ratio of [\ion{Ne}{iii}] line flux over the [\ion{O}{ii}]
flux of flux([\ion{Ne}{iii}])/flux([\ion{O}{ii}]) $< 0.07$ (2
$\sigma$). Using the spectrophotometric catalogue of local emission
line galaxies of \citet{ter91}, we found that only 5 out of a sample
of 151 of the galaxies with both [\ion{O}{ii}] and [\ion{Ne}{iii}]
lines detected ($5/151 \approx 3$\%) have such a weak Neon line. With
the estimated number of [\ion{O}{ii}] emitters (see above), we expect
that $< 1$ of our 6 emission line galaxies is an [\ion{O}{ii}]
emitter.

On the basis of these four lines of arguments, we conclude that these
6 line emitters are almost certainly \lya\ emitters at $z=5.2$.

The extracted \lya\ lines were fitted with a Gaussian function to
estimate the redshift, line flux and widths (FWHMs). In Table
\ref{zlya} the properties of the \lya\ emitters are summarised. The
IDs correspond to the object's number in the catalogue. To correct for
the instrumental broadening, the observed FWHM was deconvolved with
the resolution. The star formation rates (SFR$_\mathrm{UV}$ and
SFR$_{\mathrm{Ly}\alpha}$) were calculated from the measured UV
continuum fluxes and line fluxes in the images assuming a flat $f_\nu$
spectrum and UV flux density to SFR conversion of \citet{mad98}.

The velocities of the six confirmed \lya\ emitters cluster within a
range of 900 \kms\ in the rest-frame, while the narrow-band filter is
$\sim 3500$ \kms\ wide. The peak of the \lya\ emission of the radio
galaxy is roughly 1000 \kms\ away from the central velocity of the
emitters. This is different from other $z > 2$ radio galaxy
protoclusters, where the radio galaxy has a velocity close to the
average velocity of the \lya\ emitters
\citep{pen00a,kur04,ven02}. This could be due to \ion{H}{i} absorption
on the \lya\ emission line of the radio galaxy. It has been shown that
in radio galaxies this absorption can cause a velocity shift of the
\lya\ line up to 1000 \kms\ as compared to other UV emission lines
\citep[e.g.][]{rot97}.

Of the remaining objects covered by the mask, one is identified as a
[\ion{O}{ii}] $\lambda 3727$ emitter, also showing [\ion{Ne}{iii}] $\lambda
3869$ emission and nine were identified as [\ion{O}{iii}] $\lambda
5007$ emitters, confirmed by various lines such as [\ion{O}{iii}]
$\lambda 4959$, H$\beta$, H$\gamma$, H$\delta$, [\ion{Ne}{iii}]
$\lambda 3869$ and [\ion{O}{ii}] $\lambda 3727$. In total 11
[\ion{O}{iii}] emitters were confirmed in the field, all having a
redshift of $z \sim 0.5$. 

\begin{table*}
\begin{center}
\caption{Properties of the eight spectroscopically observed high
  priority candidates and the radio galaxy.} 
\label{zlya}
\begin{tabular}{ccccccccc}
\hline
\hline
Object & \multicolumn{2}{c}{Position} & $z$ & Flux & EW$_0$ & FWHM &
SFR$_{\mathrm{UV}}$ & SFR$_{\mathrm{Ly}\alpha}$ \\ 
{} & $\alpha_{\mathrm{J}2000}$ & $\delta_{\mathrm{J}2000}$ & {} &
\ergscm\ & \AA & \kms\ & M$_{\sun}$ yr$^{-1}$ & M$_{\sun}$ yr$^{-1}$ \\ 
\hline
 259 & 09 24 07.07 & $-$22 02 09.2 & 5.1834 $\pm$ 0.0002 & 8.8 $\pm$
 0.8 $\times 10^{-18}$ & $>$ 103 & 208 $\pm$ 39 & $<$ 3.2 & 3.9 $\pm$ 0.7 \\
 524 & 09 24 09.41 & $-$22 02 00.5 & 5.1683 $\pm$ 0.0003 & 1.2 $\pm$
 0.1 $\times 10^{-17}$ & 97$^{+866}_{-21}$ & 392 $\pm$ 25 & $9.1 \pm
 3.6$ & 10.4 $\pm$ 1.2 \\
1388 & 09 24 16.68 & $-$22 01 16.9 & 5.1772 $\pm$ 0.0003 & 4.1 $\pm$
0.5 $\times 10^{-18}$ & 59$^{+476}_{-14}$ & 295 $\pm$ 38 & $5.1 \pm
2.0$ & 3.5 $\pm$ 0.5 \\
2688 & 09 24 25.67 & $-$22 03 01.1 & 5.1731 $\pm$ 0.0003 & 3.1 $\pm$
0.4 $\times 10^{-18}$ & $>$ 88 & 167 $\pm$ 63 & $<$ 2.9 & 3.0 $\pm$ 0.6 \\
2849 & 09 24 24.30 & $-$22 02 30.9 & 5.1765 $\pm$ 0.0003 & 8.0 $\pm$
0.9 $\times 10^{-18}$ & 47$^{+785}_{-13}$ & 249 $\pm$ 32 & $6.1 \pm
2.7$ & 3.3 $\pm$ 0.7 \\
2881 & 09 24 23.88 & $-$22 03 44.8 & 5.1683 $\pm$ 0.0005 & 1.4 $\pm$
0.2 $\times 10^{-17}$ & 42$^{+45}_{-9}$ & 479 $\pm$ 28 & $13.7 \pm
3.4$ & 6.8 $\pm$ 1.1 \\
\hline
 463$^a$ & 09 24 08.48 & $-$22 00 04.0 & 0.4983 $\pm$ 0.0001 & 4.7 $\pm$
 0.7 $\times 10^{-18}$ & 339$^{+1966}_{-91}$ & $<$ 200 & -- & -- \\
 559$^a$ & 09 24 09.51 & $-$22 00 18.3 & 0.51515 $\pm$ 0.00003 & 1.3 $\pm$
 0.1 $\times 10^{-17}$ & 207$^{+160}_{-47}$ & $<$ 70 & -- & -- \\
\hline
 RG  & 09 24 19.90 & $-$22 01 42.0 & 5.1989 $\pm$ 0.0006 & 2.1 $\pm$
 0.2 $\times 10^{-17}$ & 83$^{+148}_{-14}$ & 1161 $\pm$ 55 & $11.6 \pm
 3.5$ & 11.4 $\pm$ 0.7 \\
\end{tabular}
\end{center}
$^a$ [\ion{O}{iii}] $\lambda 5007$ emitter
\end{table*}

\section{Discussion and conclusions}

The fraction of foreground contaminants in our sample is estimated to
be $2/8 \sim 25$\%. There are 6 additional unconfirmed high priority
candidate \lya\ emitters in the field. Based on the fraction of
contaminants in our sample, $\sim 4$ of those are expected to be $z
\sim 5.2$ \lya\ emitters.

Is there an overdensity of \lya\ emitters near TN J0924--2201? To
investigate this question, we have to compare the density of \lya\
emitters in our field with the density in blank fields. The largest
survey near $z \sim 5$ for \lya\ emitters is the search for \lya\
emitters at $z \sim 4.79$ in the Subaru Deep Field
\citep[SDF,][]{shi04}. This survey is comparable in depth to our
observations ($L_{\mathrm{lim}}(\mathrm{Ly}\alpha) = 3 \times 10^{42}$
erg\,s$^{-1}$ for an emitter at $z = 4.79$ with no continuum) and the
selection criteria applied to identify \lya\ emitters are very similar
to ours \citep[EW$_\lambda > 80$ \AA,][]{shi04}. In the SDF, Shimasaku
et al.\ find 51 candidate \lya\ emitters in an area of 25\arcmin\
$\times$ 45\arcmin. However, there is no spectroscopic confirmation of
these candidates. We therefore conservatively assume that all their
candidates are \lya\ emitters at $z \sim 4.8$, resulting in a number
density of \lya\ emitters in the SDF of $2.1 \pm 0.3 \times 10^{-4}$
Mpc$^{-3}$ \citep{shi04}. Excluding the radio galaxy, the density of
confirmed \lya\ emitters in our field is $5.3^{+3.2}_{-2.1} \times
10^{-4}$ Mpc$^{-3}$, which is a factor $2.5^{+1.6}_{-1.0}$ higher than
in the SDF. If the four unconfirmed candidate \lya\ emitters are
included, this factor rises to $4.2^{+2.0}_{-1.4}$. We used the data
from the SDF to estimate the chance of finding 6 or more \lya\ emitters
in within a single 6\farcm8$\times$6\farcm8 FORS2 field by counting
the number of emitters in randomly placed 6\farcm8$\times$6\farcm8
apertures. In only 7\% of the cases, more than 6 \lya\ emitters were
found. This further indicates that the TN J0924--2201 field is
overdense in \lya\ emitters.

\lya\ emitters at high redshift show large cosmic variance in their
clustering properties \citep[e.g.][]{shi04}. Various authors have
found that the distribution of \lya\ emitters on the sky and/or in
redshift space can be very inhomogeneous
\citep[e.g.][]{ouc03,fyn03,shi03,pal04,hu04}. For example, most of the
\lya\ emitters found at $z=4.86$ in the SDF are concentrated within a
large-scale structure with a radius of $\sim 6$\arcmin\ \citep[$\sim
2.5$ Mpc,][]{shi03}. It is therefore possible that the \lya\ emitters
around TN J0924--2201 in the $\sim 6$\farcm8$\times$6\farcm8 field of
view of FORS2 are located inside such a large-scale structure.

It is interesting to compare the (over)density in this field with the
protoclusters that were found around radio galaxies at $z=4.1, 3.1$
and 2.2, each with at least 20 confirmed protocluster members and
estimated masses of $\sim 10^{14}-10^{15}$ M$_{\sun}$
\citep[][Venemans et al.\ in prep]{pen00a,kur04,ven02}. In the TN
0924--2201 field objects were selected with a (\lya) line luminosity
of $> 3 \times 10^{42}$ erg\,s$^{-1}$. At $z = 4.1, 3.1$ and 2.2, this
luminosity limit corresponds to a limit of $> 1.5, 3.1$ and 7.0
$\times 10^{-17}$ \ergscm. The number of candidate (confirmed)
emitters with a line brighter than the luminosity limit in the $z =
4.1, 3.1$ and 2.2 protoclusters is 10 (10), 12 (12) and 8 (6)
respectively. This is roughly the same number of \lya\ emitters as in
the TN J0924--2201 field, which contains six confirmed and four
possible \lya\ emitters. The \lya\ emitters at $z= 5.2$ might
therefore be the bright end of a population of star forming galaxies
in a protocluster at $z = 5.2$, making it the most distant known
protocluster. Deep multi-color observations should confirm this by
detecting other populations of galaxies (e.g.\ Lyman break galaxies)
in the protocluster.

\begin{acknowledgements}
We thank the staff on Paranal, Chile for their splendid support, and
William Grenier of Andover Corporation for his help in our purchase of
the customised narrow-band filter. We also thank the referee, J.\
Fynbo, for his comments that improved this manuscript. GKM
acknowledges funding by an Academy Professorship of the Royal
Netherlands Academy of Arts and Sciences (KNAW). The work by WvB was
performed under the auspices of the U.S.\ Department of Energy,
National Nuclear Security Administration by the University of
California, Lawrence Livermore National Laboratory under contract No.\
W-7405-Eng-48. The NRAO is operated by associated universities Inc,
under cooperative agreement with the NSF. This work was supported by
the European Community Research and Training Network ``The Physics of
the Intergalactic Medium''.
\end{acknowledgements}

\end{document}